\documentclass[12pt,preprint]{aastex}

\def\a4{\hsize 17.0cm \vsize 25.cm}
\newcommand{\der}[2]  { \frac{{\rm d}#1}{{\rm d}#2} }

\shorttitle{On the Extreme Positive Feedback from SSCs}
\shortauthors{Tenorio-Tagle et al.}

\begin{document}

\title{On the Extreme Positive Feedback Star-Forming Mode from
       Massive and Compact Superstar Clusters}

\author{
Guillermo Tenorio-Tagle, Sergiy Silich, Ary Rodr\'\i{}guez Gonz\'alez}
\affil{Instituto Nacional de Astrof\'\i sica Optica y
Electr\'onica, AP 51, 72000 Puebla, M\'exico; gtt@inaoep.mx}

\and

\author{Casiana Mu\~noz-Tu\~n\'on}
\affil{Instituto de Astrof\'{\i}sica de Canarias, E 38200 La
Laguna, Tenerife, Spain; cmt@ll.iac.es}

\begin{abstract}
The force of gravity acting within the volume occupied by young,
compact and massive superstar clusters, is here shown to drive in 
situ all the matter deposited by winds and supernovae into several 
generations of star formation. These events are promoted by radiative 
cooling which drains the thermal energy of the ejected gas causing 
its accumulation to then rapidly exceed the gravitational instability 
criterion. A detailed account of the integrated ionizing radiation and 
mechanical luminosity as a function of time is here shown to lead to a 
new stationary solution. In this, the mass deposition rate $\dot M$, 
instead of causing a wind as in the adiabatic solution, turns into a 
positive feedback star-forming mode equal to the star formation rate. 
Some of the implications of this extreme positive feedback mode are 
discussed.
\end{abstract}


\section{Introduction}

The discovery by HST and large ground-based telescopes, of a
large population of superstar clusters (SSCs) in a large variety of
galaxies (see Ho 1997,  Mart\'\i{}n Hern\'andez et al. 2005, and 
Melo et al. 2005), has driven the community to speculate about 
the impact that these new units of star formation may have on the ISM 
of their host galaxies. SSCs with a mass, M$_{SC}$, ranging from a 
few $\times 10^5$ M$_\odot$ to up to 6 $\times 10^7$ M$_\odot$ 
(see Walcher et al. 2004; Pasquali et al. 2004) all within a small 
volume of radius 3 to 10 pc, are indeed some of the most energetic 
entities within galaxies. Their large UV photon output and mechanical 
energy are now believed to be the largest negative feedback agents 
in starburst galaxies, leading not only to a large-scale  structuring
of the ISM and to limit star formation, but also to be the agents
capable of establishing as in M82 a supergalactic wind, thereby 
removing processed material from galaxies and causing the
contamination of the IGM (Tenorio-Tagle et al. 2003).

In the present series of papers however, we have shown that there is 
a threshold limit that massive and energetic compact clusters may 
cross to find out that radiative cooling inhibits their  stationary 
outflow condition (Silich et al., 2004) and then the matter ejected 
by the stellar sources, unable to escape, is to accumulate within the  star 
cluster volume (Tenorio-Tagle et al. 2005). The facts in such cases
are that radiative cooling drastically diminishes the sound speed and
the pressure gradient within the thermalized ejected matter,
inhibiting the possibility of a wind. Radiative cooling upsets the 
stationary condition in which the deposited matter ($\dot M $) has to 
equal, at all times, the amount of matter streaming out of the SC volume, 
$\dot M = 2 \mathcal {L}_{SC}/v_\infty^2 = 4 \pi R_{SC}^2 \rho_{SC} c_{SC}$,
where $\mathcal {L}_{SC}$ is the amount of energy injected
by stellar winds and SNe, $c_{SC}$ is the local sound speed at the 
SC surface and $v_\infty$ is the resultant wind terminal speed in 
the absence of radiative cooling. As soon as this happens, the ejecta 
begins to accumulate, promoting larger densities and an even faster cooling
within the SSC volume.

The location of the threshold line  in the $\mathcal {L}_{SC}$ {\it
vs} size ($R_{SC}$) diagram (see Figure 1), the line that defines
whether or not a wind is inhibited,  depends on several variables. It 
depends on the size of the star-forming region ($R_{SC}$), the 
metallicity of the ejected gas, which has a strong impact on the 
cooling curve (see paper III). It also depends on the 
assumed $\mathcal {L}_{SC}/\dot M$ or adiabatic terminal speed 
($v_{\infty}$) of the wind. The latter is also bound to the usual 
assumption that the energy deposited by SN is always $10^{51}$ erg but 
the mass of potential SNe within the cluster ranges from, say, 
100 M$_\odot$ to 8 M$_\odot$ and so the injection speed (similar to 
$v_{\infty}$) and the deposited amount of matter are also functions 
of time. Another factor that strongly affects the location of the 
threshold line is the thermalization efficiency ($\epsilon$) which 
simply defines the fraction of the mechanical energy that can be 
evenly spread within the cluster volume. Estimates of $\epsilon$ by 
several authors lead to values between 1 (Chevalier \& Clegg; 1985) to 0.03 
(see Melioli \& Del Pino 2004 and references therein) and depends 
simply on the proximity of the sources undergoing winds and SNe, which 
through radiation may reduce the amount of energy available after 
thermalization. We have shown that there are three different types of 
solutions: SSCs far away from the threshold line (low mass,
low energy clusters) undergo a quasi-adiabatic evolution well 
described by the Chevalier \& Clegg (1985) and Cant{\'o} et al., (2000)
solution. More energetic clusters 
are to have strongly radiative winds. Cooling hardly affects their 
velocity ($v_w \sim v_{\infty}$) and density distribution 
($\rho_w \propto r^{-2}$), but their temperature instead of falling 
as $r^{-4/3}$, it falls rapidly to $T_w \sim 10^4$ K close to the 
SC boundary and the more so, the closer they are to the threshold 
line. The strongly radiative winds  around such clusters lead, 
compared to the adiabatic solution, to very much reduced X-ray 
envelope sizes. The third solution is for clusters  above the threshold line.
These would have their winds inhibited and as shown below, this  
turns them into very efficient positive feedback star-forming agents.

Here we follow the evolution of a massive and compact cluster and 
show the fate of the mass reinserted back, through winds and SNe, 
into the SC volume. Section 2 and 3 give a full description of SSC as 
positive feedback agents and section 4 summarizes our conclusions.

\section{Feedback from massive and compact SSC}

Here we follow the evolution of a stellar cluster, assumed to
be the result of an instantaneous or coeval burst of  star formation. 
The cluster has an initial total mass in stars ($M_{SC}$)
equal to 10$^7$ M$_\odot$ within an $R_{SC}$ = 5 pc.  We assume a 
standard Salpeter IMF and an upper and lower mass limits equal to
100  M$_\odot$ and 1  M$_\odot$, respectively. The mechanical energy
of the cluster $\mathcal {L}_{SC} \sim 10^{41} $erg s$^{-1}$ places
our example cluster in the zone where stationary superwinds are 
inhibited (see Figure 1) and thus the mass returned by the stars ($\dot M $)
begins immediately to accumulate within the star cluster volume, 
favoring an even faster cooling. Figure 2 shows how the density of 
the accumulating gas ($\rho_{ac} = 3 \dot M t/ 4 \pi R_{SC}^3$; where  
$t$ the evolution time), grows as a function of time  within the SSC volume. 

Following this trend, clearly there is a moment when the density of
the accumulating gas will exceed the gravitational instability criteria 
\begin{equation}
\label{jeans}
\rho_J = \frac{\pi \gamma k }{4 G \mu } T R_{sc}^{-2}\sim 2.3 
\times 10^{-20} \left(\frac{T}{100K}\right) \left(R_{SC1}\right)^{-2} 
\; g \, cm^{-3}
\end{equation}
where $\rho_J$ is the Jeans density and $k, G, \mu$ and $\gamma$  
are the Boltzmann constant, gravitational constant, mean mass per 
particle and the ratio of specific heats (5/3), respectively
and  $R_{SC1}$ is the SSC radius in pc units. At that moment, 
when $\rho_{ac} = \rho_J$ collapse will inevitably proceed.

The initial ample supply of UV photons ($N^0$ = 10$^{54}$ 
photons s$^{-1}$, see Figure 3a), exceeds at first the number of 
recombinations within the volume occupied by the reinserted gas and 
the resultant HII region,  given the large metallicities of the
ejecta, is here assumed to rapidly approach an equilibrium temperature 
$T_{HII} \sim$ a few $ 10^3$ K. At these temperatures the sound
speed ($<$ 10 km s$^{-1}$) remains well below the escape speed 
($V_{esc} = (2 G M_{SC} / R_{S})^{1/2} \sim 130$ km s$^{-1}$) and the
reinserted gas would inevitably continue to accumulate to rapidly 
(within $1.5 \times  10^6$ yr) reach the value of the Jeans density 
for a gas at say, $T_{HII}$ = 5000 K, and collapse into a new stellar 
generation. The event gives rise to a new phase of matter
accumulation, which once more will rapidly approach $\rho_J$ 
(for $T$= 5000 K) and undergo collapse within a free-fall time, of 
the order of 10$^5$ yr, while transforming  $\sim 2 \times 10^5$ M$_\odot$
into stars. All stellar generations resultant from mass accumulation 
within the SSC volume have here been assumed to also acquire a
Salpeter IMF with similar upper and lower mass limits as those imposed 
to the main superstar cluster, and their resultant properties 
(mechanical energy and UV photon output) have been added to those 
produced by the main cluster.

A few, almost identical, stellar generations are expected  from the 
accumulation process (solid rising lines in Figure 2), every time 
that the accumulated gas density $\rho_{ac}$ reaches $\rho_J(5000K)$ 
(dashed line in Figure 2). The situation changes slightly when the 
number of ionizing photons ($N^0$), despite the added contribution  
of secondary stellar generations, becomes insufficient to fully ionize 
the accumulated matter within the star cluster volume. This is due to the 
evolution of the main cluster, whose UV photon output begins to fall 
as $t^{-5}$ after $\sim$ 3.5 Myr (dotted line in Figure 3a). 
Figure 2 shows $\rho_{HII}$ (thin solid line), the maximum  density
within the accumulating volume that  can be supported fully ionized by 
the UV radiation produced by the evolving cluster
($\rho_{HII} = (3 N^0 \mu^2 / 4 \pi R_{SC}^3 \beta)^{1/2}$; where 
$\mu = 1.4 m_H$ and $\beta$, the recombination coefficient
to all levels above the ground level = 2.59 $\times 10^{-13}$ 
cm$^{-3}$ s$^{-1}$). During the accumulation process, once $\rho_{ac}$ 
exceeds $\rho_{HII}$, the ionized volume begins to shrink to end up 
as a collection  of ultra compact HII regions around the most massive 
stars left within the cluster, while the bulk of the ejected material, 
now recombined, continues to cool, approaching rapidly (within a 
time-scale of less than 1000 yr) a temperature $\sim$ 100 K. 
Matter is at all times uniformly replenished 
within the whole SSC volume, and thus the gas density presents an 
almost  uniform value. However, the accumulating gas  now has two 
different temperatures ($T_{HII}$ and 100 K) and as $\rho_{ac}$ 
grows and the fraction of the ionized volume 
($f_{HII} = 3N^0\mu^2 / 4\pi R_{SC}^3 \beta \rho_{ac}^2$) 
shrinks, the size of cold  condensations (at 100 K) able to  
become gravitationally unstable and their free-fall time also 
become smaller.  
 
The drop in the number of ionizing photons (Figure 3a) and the
consequent growth of the neutral volume (Figure 3b) lead then to a  
second important condition in which the  characteristic accumulation time 
$\tau_{ac} = \frac{4}{3} \pi \rho_{gas}   R_{SC}^3 (1 - f_{HII})/ {\dot M}$ 
becomes equal to the free-fall time 
$t_{ff}= (3 \pi / 32 G \rho_{gas})^{1/2}$ .
This condition defines $\rho_{gas}$, the density above the Jeans 
instability limit for a neutral condensation at 100 K ($\rho_J$(100K)): 
\begin{equation}
\label{eq.3} 
\rho_{gas} = \left[\frac{27 {\dot M}^2}{512 \pi G \left(1 -f_{HII}
             \right)^2} \right]^{1/3} 
R_{SC}^{-2} = \frac{1.15 \times 10^{-18}}{(1 - f_{HII})^{2/3}} 
              \left(\frac{\dot M}{1M_\odot yr^{-1}} 
              \right)^{2/3} \left(\frac{R_{SC}}{1pc}\right)^{-2} ,
\end{equation} 
and thus once $\rho_{ac}$ (see Figure 2) becomes equal to
$\rho_{gas}$, a new stationary solution becomes possible.

Everything happens  very rapidly, compared to the evolution time-scale  
of the parental cluster ($\sim$ 40 Myr), and almost at the same time. 
The ejected matter is thermalized within the SSC volume and
immediately begins to cool. At  the same time that it accumulates 
making cooling even faster. This allows it to rapidly reach the 
required $\rho_{gas}$ value, above the Jeans instability limit, 
that warrants its collapse in a similar time-scale, while the 
collapsing material is replenished by the newly ejected matter.
When this happens, the mass deposition rate from the cluster becomes equal to
the rate of star formation. Gravitational collapse and star formation 
within the star cluster volume and with the matter injected by all 
sources, drive in this way a new stationary condition through  a new 
era of quasi-continuous star formation in which $\dot M$ is equal now 
to the star formation rate (SFR).

\section{The quasi-continuous star formation era}

All new generations of stars are due to deposit their ejecta from 
winds and SNe during $\approx 40$ Myr, the expected duration of the 
type II SN phase in  coeval star clusters (Leitherer \& Heckman 1995), 
and at a rate that depends only on the mass of the stellar
generation. Re-processing leads to an escalating  
$\dot M$, as this is directly proportional to the total mass in stars
\begin{equation}
M_{stars}(t) = M_{sc} + \int_{t_C}^{t} SFR(t) \rm{d}t  
\end{equation}  
created within  the star cluster volume; where $t_{C}$ marks the start 
of the quasi-continuous star-forming phase and $M_{SC}$ accounts for 
the initial mass in stars and for all generations prior to the 
quasi-continuous star formation phase (see Figure 2). This  promotes 
a larger star formation rate, and a faster fulfillment of the collapse 
conditions ($\rho_{gas} \geq \rho_J$, and $\tau_{ac} = t_{ff}$). 
Under these conditions the SFR is:
\begin{equation}
\label{sf.a} 
SFR(t) = \der{M_{stars}}{t} = \delta M_{stars}(t) .
\end{equation}  
This equation has an exponential solution:
\begin{equation}
\label{sf.b} 
M_{stars}(t) = M_{sc} \exp{[(t - t_C) / \tau]} ,
\end{equation}  
where  the characteristic time $\tau = \delta^{-1}$. Equation  
(\ref{sf.b}) remains  valid during the supernovae phase of the
parental cluster. Within this time interval, the continuous
re-processing of the ejected material drives, as shown in Figure 4, 
the star formation rate to grow exponentially:
\begin{equation}
\label{sf.c} 
SFR(t) = \der{M_{stars}}{t} = \frac{M_{sc}}{\tau} 
      \exp{[(t - t_C) / \tau]} .
\end{equation}  

Note that the characteristic time $\tau$ is defined by the coeval star cluster
model and is applicable to both  the initial cluster 
and also to every new generation of stars formed from the re-ejected 
material. If one assumes that the energy and mass deposition rates remain 
constant during a star cluster life time and that $v_{\infty}$, 
the resultant terminal speed in the absence of radiative cooling,
equals 1500 km s$^{-1}$ for all clusters, upon normalizing to the 
energy deposition rate adopted for the parental 10$^7$ M$_\odot$ 
cluster ($\mathcal {L}_i = 10^{41} M_i/10^7M_\odot$ erg s$^{-1}$),  then
$\tau = M_{i} / {\dot M}_i = M_{i} v_{\infty }^2 / (2 \mathcal {L}_i) 
 = 7.5 \times 10^7$ yr, where $M_i$ is the mass in each stellar 
generation, including the initial or parent cluster. The total mass 
in stars born within the star cluster volume, over the characteristic 
time $\tau \approx 40$ Myr, is given by equation  (\ref{sf.b}).

\section{Conclusions}

We have here shown that massive and compact coeval clusters instead 
of driving a superwind able to disperse the surrounding ISM
and even channel  its way into the IGM, events that have make them 
been regarded as negative feedback agents, they 
are in fact extreme examples of positive star formation feedback. 
 
Massive and compact coeval clusters appear in the $\mathcal 
{L}_{SC}$ {\it vs} size diagram above the threshold line,
in the region where radiative cooling inhibits the development of 
stationary superwinds. We have here shown that in such cases the 
matter reinserted, through stellar winds and supernovae, is unable 
to escape and that after a short phase of matter accumulation,  a 
new stationary solution in which $\dot M$ becomes equal to the SFR 
is rapidly met. A positive feedback condition in which new stellar 
generations result in situ, from the collapse of the matter reinserted  
by the sources evolving within the star cluster volume.

The secondary star formation process while causing a faster mass 
deposition rate, drives the SFR to grow from 0.1 to 0.25 
M$_\odot$ yr$^{-1}$ over the parent cluster supernova phase ($\sim $ 40 Myr).
The continuous re-processing of the ejected material leads effectively 
to a continuous transformation of the high mass stars into a low mass 
($\leq$ 8 M$_\odot$) population, keeping constant the total mass of 
the stellar component. In this way massive clusters may survive
as gravitationally bound systems (see discussion in Dale et al. 
2005 and Mart\'\i{}n-Hern\'andez et al. 2005 and references therein).

A central issue, regarding ISM studies, is the fact that the more 
massive and compact clusters (as those detected by HST and 
large ground-based telescopes), are unable to generate superwinds and  
shed their metals into the ISM or the IGM. Their evolution leads to 
many stellar generations and thus to a mixture of stellar populations, 
all contaminated by the products from former stellar generations.
Exacerbated episodes of star formation that leave no trace of their 
evolution in the ISM. Clearly, if the most massive star formation 
events do not contribute to the chemical evolution of their parent 
galaxy, then several issues, such as the  mass-metallicity relationship 
between galaxies, ought to be revised.

\acknowledgments 
We thank our anonymous referee for comments and suggestions.
This study has been supported by CONACYT - M\'exico, research grant 
47534-F and AYA2004-08260-CO3-O1 from the Spanish Consejo Superior de
Investigaciones Cient\'\i{}ficas.

\newpage

\figcaption[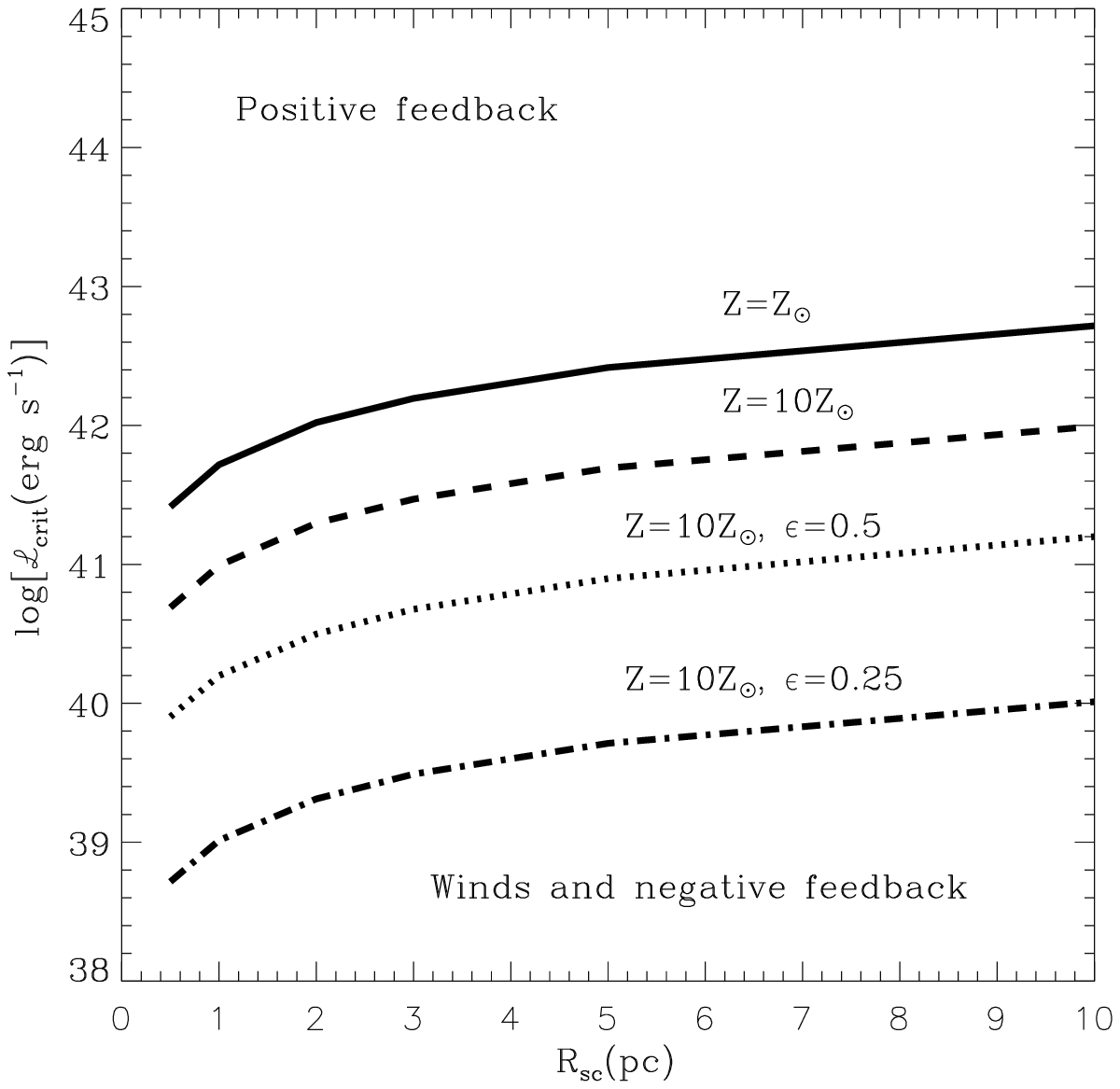]
{The threshold line. The location of the threshold line, the line that 
divides the $\mathcal {L}_{SC}$ {\it vs} size ($R_{SC}$) diagram into 
two distinct areas. There the matter deposited within the SSC volume, 
through winds and SNe, ends up either streaming away as a stationary 
(adiabatic or radiative) wind (below the line) or it accumulates to 
end up steadily being driven into new episodes of star formation
(above the line). Two different locations of the threshold line, for 
an assumed full thermalization efficiency ($\epsilon$ = 1), are
display for an ejecta metallicity value equal to solar and ten times solar. 
The change in location  for different values of $\epsilon$ is also indicated.}

\figcaption[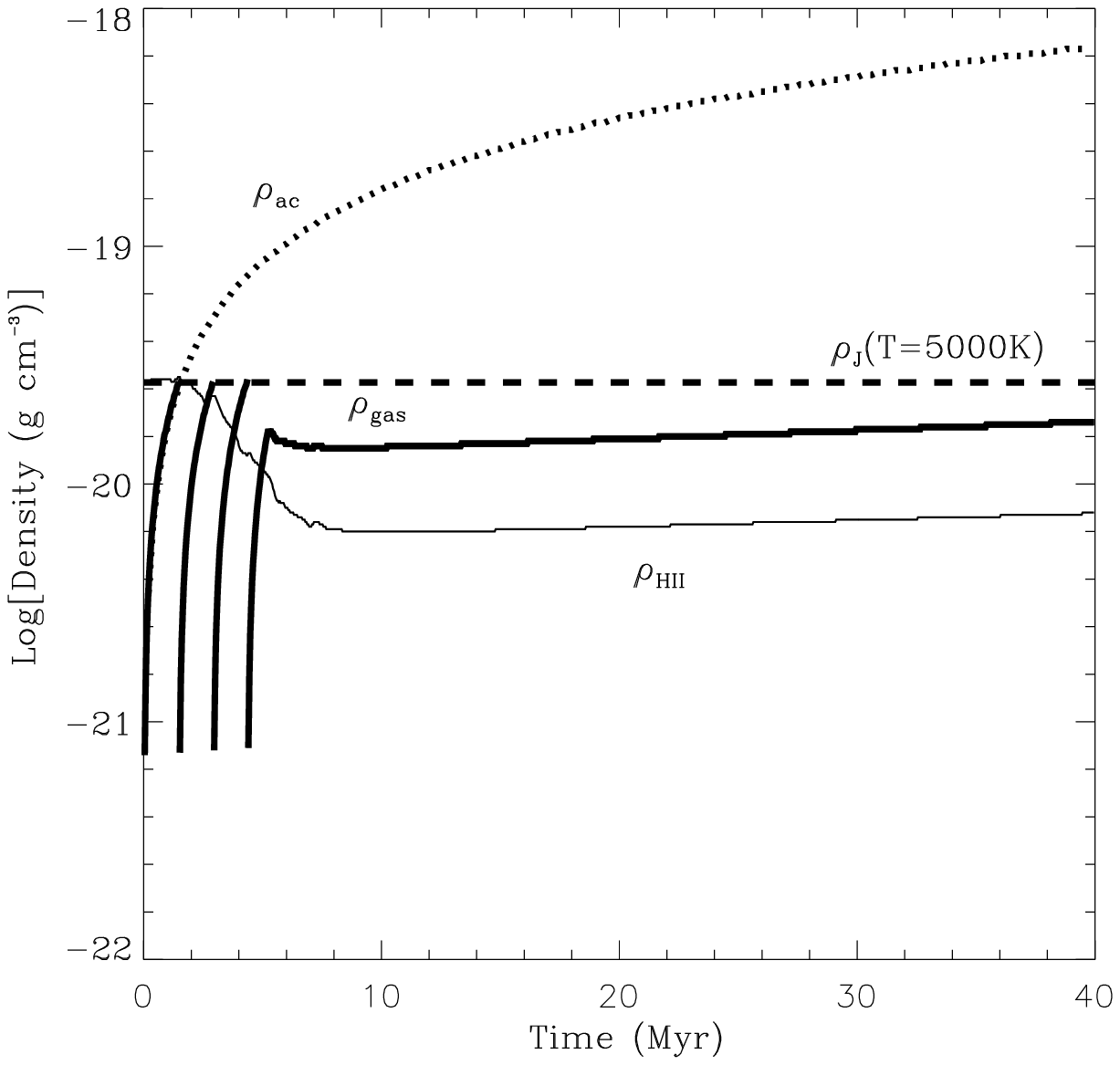]
{Positive star-forming feedback. The figure shows the rapid growth of
the reinserted gas density within the SSC volume (rising solid lines), 
interrupted as it reaches the gravitational instability criterium ($\rho_J$), 
leading to a new stellar generation and to a new phase of matter
accumulation. The situation changes slightly as the available photon
flux becomes unable to sustain all the SSC volume fully ionized,
allowing for recombination and further cooling of the recombined matter,
and thus to a much lower limit of the Jeans instability criterium.
This allows the accumulation time to become comparable to the free-fall time, 
establishing a new stationary condition in which $\dot M$ = SFR. The 
density value ($\rho_{gas}$) required for this condition is also
indicated in the figure as well as the maximum density value ($\rho_{HII}$)
that can be supported fully ionized within the SSC volume, with the 
available UV photon flux.} 

\figcaption[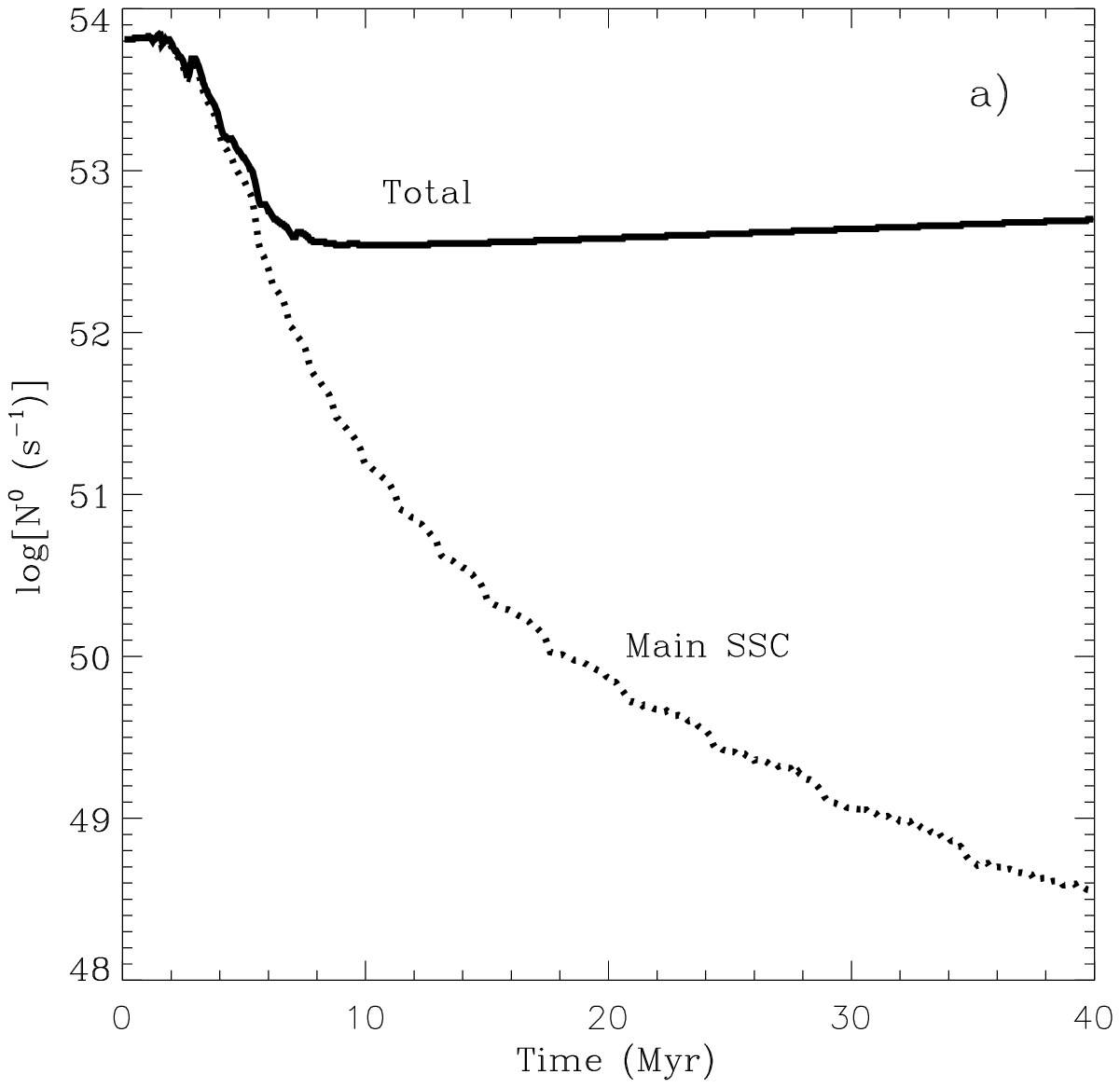, 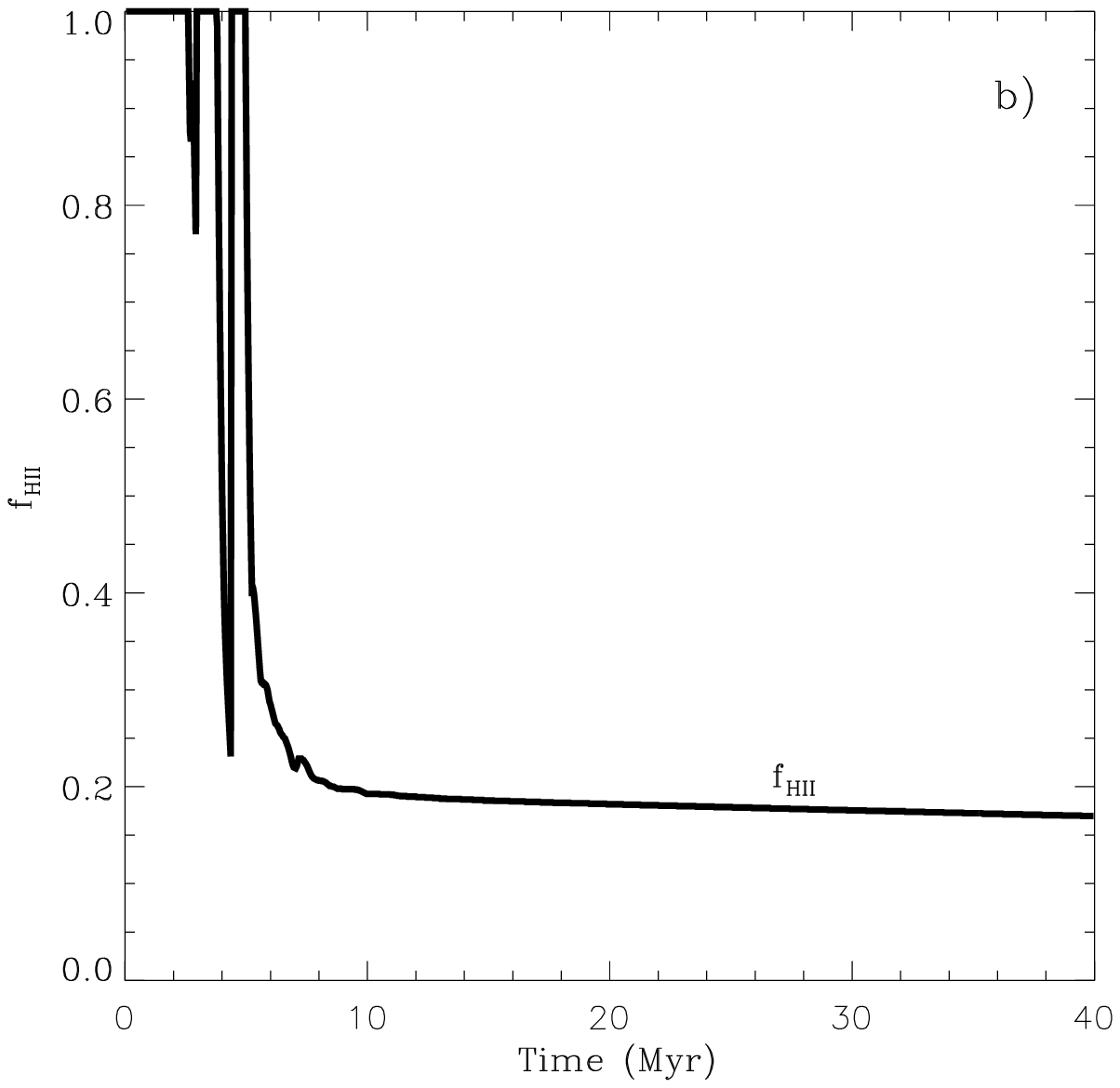]
{Figure (a) show the evolution of the UV photon flux produced by the
main star cluster (dotted line) and the total flux resultant from the 
contributions from all further stellar generations (solid line). b) 
displays $f_{HII}$, the fraction of the SSC volume that can be kept 
fully ionized with the available UV stellar photon output.}  

\figcaption[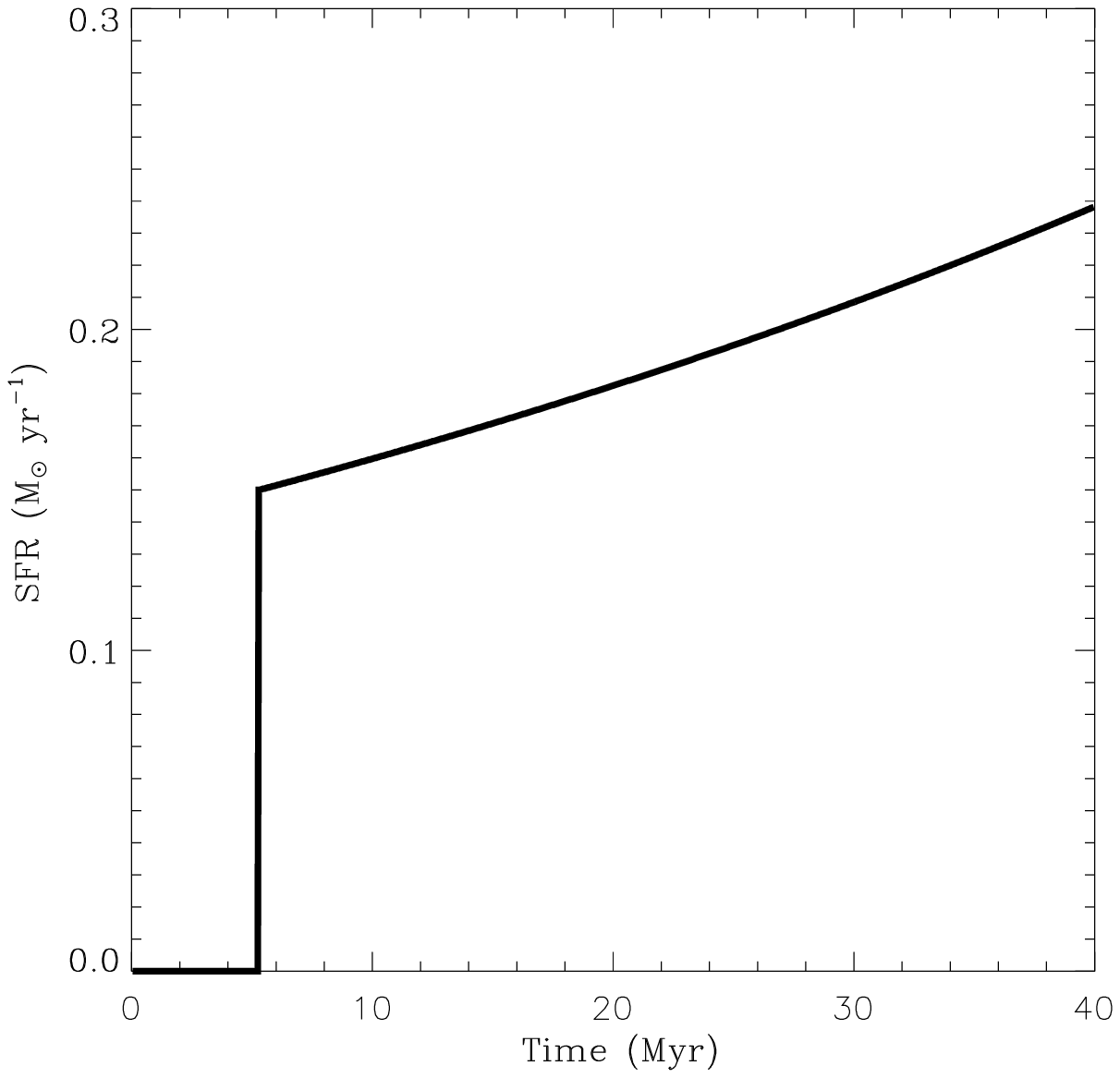]
{Positive star-forming feedback. The evolution of the star formation 
rate throughout the life-time of the main cluster.}   

\newpage

\begin{figure}[htbp]
\plotone{fig1.ps}
\end{figure}

\newpage

\begin{figure}[htbp]
\plotone{fig2.ps}
\end{figure}

\newpage

\begin{figure}[htbp]
\plotone{fig3a.ps}
%
\plotone{fig3b.ps}
\end{figure}

\newpage

\begin{figure}[htbp]
\plotone{fig4.ps}
\end{figure}




\end{document}